\documentclass[twocolumn]{bmcart}

\usepackage{amsthm,amsmath}
\usepackage{subcaption}
\usepackage{threeparttable}
\usepackage{adjustbox}
\usepackage{multirow}
\usepackage[figuresright]{rotating}
\usepackage[numbers,sort&compress]{natbib}
\usepackage[utf8]{inputenc} 
\usepackage[ruled,linesnumbered]{algorithm2e}

\startlocaldefs
\endlocaldefs

\begin{document}

\begin{frontmatter}

\begin{fmbox}
\dochead{Research}


\title{Towards the Transferable Audio Adversarial Attack via Ensemble Methods}


\author[
    addressref={aff1,aff2},                   
   email={guo\_feng@mail.sdu.edu.cn }
]{\inits{GF}\fnm{Feng} \snm{Guo}}
\author[
  addressref={aff1,aff2},
  email={sun\_zheng@mail.sdu.edu.cn }
]{\inits{SZ}\fnm{Zheng} \snm{Sun}}
\author[
  addressref={aff1,aff2},
  corref={aff1,aff2},    
  email={chenyuxuan@sdu.edu.cn}
]{\inits{CYX}\fnm{Yuxuan} \snm{Chen}}
\author[
  addressref={aff1,aff2},
  email={julei@sdu.edu.cn}
]{\inits{JL}\fnm{Lei} \snm{Ju}}

\address[id=aff1]{
  \orgdiv{School of Cyber Science and Technology},             
  \orgname{Shandong University},          
  \city{Qingdao},                              
  \cny{China}                                    
}
\address[id=aff2]{
  \orgdiv{Quancheng Laboratory},         
  \orgname{QCL},          
  \city{Jinan},                              
  \cny{China}                                    
}



\end{fmbox}


\begin{abstractbox}

\begin{abstract} 
In recent years, deep learning (DL) models have achieved significant progress in many domains, such as autonomous driving, facial recognition, and speech recognition. However, the vulnerability of deep learning models to adversarial attacks has raised serious concerns in the community because of their insufficient robustness and generalization. Also, transferable attacks have become a prominent method for black-box attacks. In this work, we explore the potential factors that impact adversarial examples (AEs) transferability in DL-based speech recognition. We also discuss the vulnerability of different DL systems and the irregular nature of decision boundaries. Our results show a remarkable difference in the transferability of AEs between speech and images, with the data relevance being low in images but opposite in speech recognition. Motivated by dropout-based ensemble approaches, we propose random gradient ensembles and dynamic gradient-weighted ensembles, and we evaluate the impact of ensembles on the transferability of AEs. The results show that the AEs created by both approaches are valid for transfer to the black box API.
\end{abstract}


\begin{keyword}
\kwd{Adversarial Attacks}
\kwd{Dynamic Gradient Weighting}
\kwd{Transferability}
\kwd{Ensemble Methods}
\end{keyword}


\end{abstractbox}
%

\end{frontmatter}



\section*{Introduction}
Recent studies \cite{11,12,13,14,15,16} have shown that tasks based on deep learning, such as image recognition and speech recognition, are vulnerable to adversarial attacks. Szegedy et al. \cite{1} first introduced the concept of adversarial attacks and adversarial examples (AEs). AEs are constructed by deliberately injecting carefully crafted subtle perturbations into the input. These adversarial examples may pose a barrier to the development of deep neural networks (DNNs), such as in facial recognition, and intelligent homes.

Adversarial attacks can be categorized as white and black-box attacks based on the attacker's knowledge of the target's model. In white-box attacks, the attacker can access details about the target model, including its structure, parameters, and even the training dataset, which allows the attacker to construct adversarial samples using techniques such as backpropagation. On the other hand, in black-box attacks, the attacker can only send input data to the target model and receive prediction outcomes without knowledge of the model's internal workings. This type of attack is particularly challenging and realistic, especially for commercial API models. Consequently, black-box attacks have garnered considerable attention in recent times.

Although some deep learning tasks can achieve an accuracy rate of 99.99\% on the training dataset, the precision in the real world may suffer heavily from the model's poor robustness and generalization ability. Adversaries have already targeted these vulnerabilities to launch attacks against selected black-box models. Several studies by academic and corporate researchers have shown that adversarial examples designed to attack specific deep learning (DL) models can also attack other DL models, referred to as “transfer attacks.” This implies that adversaries can create a substitute target model based on some publicly accessible deep learning models to tailor the perturbation and generation algorithm of AEs based on the target model's query feedback to attack the target model. This method has been widely adopted for attacks on black-box models (such as commercial black-box APIs). Many studies \cite{4,6,7,10,15} have proposed different algorithms to improve the transfer attack capability and provided relevant experimental evidence.

The majority of research on transfer attacks focuses on image classification. The robustness and generalizability of speech recognition models have not received sufficient attention. Some work \cite{39,49,40,35} has achieved positive results in attacking black-box APIs, but there is limited evidence identifying the factors strongly connected to the transferability of AEs in speech recognition. The approach to understanding speech transfer attacks is mostly to contrast the studies on images. However, speech recognition is a more complex issue than image recognition, with many technical differences that are useful for images but not for voice. There are many unique properties in adversarial attacks on voice.

In this study, we conducted a thorough analysis of the underlying factors that impact the transferability of adversarial examples (AEs). Specifically, our research entailed numerous experimental studies and data analyses. We delved into the role of noise, scale invariance, and other factors that influence the transferability of AEs and provided possible explanations for the observed phenomena. In addition, we attempted to delineate the geometric properties of DL models and uncovered the volatility of the decision boundary of the model. 

Furthermore, we observed that the application of the dropout self-ensemble technique could enhance the transferability of audio AEs. Based on this insight, we present two strategies for creating transferable AEs by combining multiple models, i.e., the random gradient ensemble and the dynamic gradient weighting ensemble. These approaches aim to optimize the transferability of AEs across different models, thereby enabling the creation of AEs that can transfer across diverse systems.

Contributions. The contributions of this paper are as follows:

\quad\noindent$\bullet$We observed notable differences in the transferability of AEs between images and voice. The relationship between pixels in images is low, but data context connections are critical in speech recognition. We first attempt to portray the geometric property of speech DL models, where we believe that the decision boundary of the speech recognition model is not as smooth and continuous as in the images, but rather irregular and unstable.

\quad \noindent$\bullet$Our experiments reveal the significant influence of noise, muted frames, scale invariance, dropout, and other factors on AEs and suggest that different factors have different degrees of contribution to transfer rate and perfectibility.

\quad \noindent$\bullet$Motivated by the success of dropout-based model ensembles, we propose random gradient ensembling and dynamic gradient weighting ensembling to generate the AEs. Our experiments suggest that the AEs generated by both proposed strategies can be implemented at a black-box application programming interface (API). Furthermore, we find that a p-value of 0.5 achieves optimal conditions for the generation and transferability of AEs.

 \quad \noindent$\bullet$We release the source code for our research at: xxx

\section*{Background}
Here, we will introduce relevant research on adversarial attacks and increasing transferability. We analyzed and summarized the fundamental conclusions on transferability.

\subsection*{\textbf{Adversarial attacks and adversarial examples (AEs)}}
Here, we concentrate on adversarial tasks. In a setup like this, the DNN network is represented as $f$, and $f:\boldsymbol{X}\to \boldsymbol{C}$ represents the given input $x(x\in \boldsymbol{X})$ is mapped to one of a set of classes $\boldsymbol{C}$, where $f(x)=c\in \boldsymbol{C}$. The DNN model is vulnerable to adversarial input attacks, which forces the DNN model to misjudge. Attacks on DNNs can be classified as targeted and untargeted. Here, we will focus on the setting of targeted attacks. Specifically, adversarial examples ${x^*}$ are normally generated by slightly modifying $x$ and ${x^*} = x + \delta $. The solve of $\delta$ can be converted to a min-optimization problem, i.e., $arg$ $min$  $\mathcal L(f(x + \delta ),c^*)$. The adversary's goal is to force $f$ to misclassify ${x^*}$ as the target ${c^*}$, i.e., $f({x^*}) = {c^*} , {c^*} \ne c $. To ensure that ${x^*}$ is acoustically similar to $x$, the perturbation needs to be restricted to a limited range $g({x^*} - x) \le \varepsilon $, where the $g$ is a measurement function of the auditory difference. 

In the audio adversarial attacks, Carlini \& Wagner \cite{4} were the first to successfully attack the Deepspeech model in 2018. At the time, Deepspeech was the most popular open-source neural network-based end-to-end speech recognition model. In that same year, Yuan et al. \cite{7} also achieved a successful attack on the Kaldi ASpIRE Chain Model\footnote{Kaldi.http://kaldi-asr.org.}. Kaldi is a voice recognition system that is widely applied by Microsoft, Xiaomi, and other companies. The following year, Qin et al. \cite{5} used an auditory masking model to attack Lingvo \footnote{https://github.com/tensorflow/lingvo.} and produced more robust AEs that can be replayed in the physical world. Lingvo is a speech recognition system developed by Google that adopts deep learning to recognize the voice and convert it into text in a highly precise and calculating efficiency. Chen et al. \cite{7} successfully attacked the APIs of Facebook, Bing, Google, AWS, and other commercial entities by a local substitute model. Zheng and Jiang \cite{40} successfully attacked voice recognition API interfaces from companies such as iFlytek, Alibaba, and Tencent by the coevolutionary algorithm. These attacks achieved nearer to 100\% success rates against commercial APIs and obtained exciting results.

Subsequently, the research on attacks against open-source deep learning models became almost static. Instead, they are shifting their attention towards attacks on commercial speech recognition APIs. The researchers doubt that these commercial APIs may leverage DL models, and hence, might be susceptible to the same vulnerabilities as open-source deep learning models.

\subsection*{\textbf{Transferability of AEs}}
The transferability of AEs is an important attribute. It means that one AE can attack different models. In other words, an AE is not just for fooling one specific mod, but can also deceive other models. From a geometric perspective, paper \cite{3} shows that the decision boundaries between different models are very similar, which provides an explanation for why AE can transfer between different models. The study quantitatively analyses the different transferability of optimization-based and gradient-based attacks, showing that the gradient-based attacks have a higher transferability. Inspired by data augmentation strategies, paper \cite{4} proposes to increase the transferability of adversarial attacks through input diversity. This approach reduces the overfitting risk of AEs, and subsequent works have proposed different input transformations to further mitigate overfitting, such as \cite{5,6,7}. Some people think that the more linear the decision boundary in the last layer, the better the transferability of AEs in DNN, as suggested by \cite{8}.  Paper \cite{9} proposes perturbing the feature space instead of the input to improve transferability and employs attention mechanisms to achieve this. Finally, \cite{10} introduces Adversarial Distribution Enhancement to tune the adversarial sample distribution and improve transferability. The paper shows that the transferability of adversarial samples is affected by various factors and suggests effective methods to enhance transferability. Future research should focus on developing more robust and generalizable methods to defend against adversarial attacks.

In another study, \cite{58} analyzed the transferability of AEs from different models in quiet and noisy settings. They identified a potential connection between transferability and additive noise and proposed an approach that injects noise during the gradient ascent process to enhance transferability. On the other hand, \cite{59} proposed an algorithm to generate universal adversarial perturbations with high transferability. The perturbation is updated using gradient information. In a parallel study, \cite{60} introduced the Houdini attack method, which exploits the difficulty of human perception of AE. The results show that Houdini has good transferability and concealment, and can successfully fool even unknown models.

The highly non-linear and complex nature of speech signals leads to the high variability and uncertainty in their features in adversarial speech recognition attacks. Simultaneously, the transmission environment of voice signals is very complicated, with noise, room acoustics, speaker variability, and so on. Therefore, adversaries can generate highly transferable adversarial samples by these features, successfully fooling multiple models.

Research has highlighted that adversarial examples do not appear randomly in small spaces but rather in large, continuous subspaces. The dimensionality of these subspaces is a critical factor in the issue of transferability since higher dimensionality increases the likelihood that subspaces of different models overlap. There is evidence that these perturbations affect specific features that different models rely on to make decisions. Thus, AEs may be transferable between different models. Additionally, models with similar structures, parameter weights, or feature extraction methods may share common points, facilitating the transfer of AEs between them. Similarly, if the source and target models have similar structures and parameters, it may be easier to transfer adversarial examples between these two models, as shown in those studies. 

\subsection*{\textbf{Properties of transferability}}
Based on the above analysis and related studies \cite{1,2,3,4,5,6,7,8,9,10,11,12,13,14,15,16}, we summarized some properties of the transferability.

Given models A, B, and C with different network structures but the same training dataset, where the number of parameters follows the relationship $A \gg B \approx C$, the following properties are observed:

\quad\noindent$\bullet$The transferability of AEs greatly relies on the model architecture. Specifically, AEs produced in model A will have worse transferability than those in model B. While larger and deeper models tend to have better performance, they may also have reduced transferability. Smaller models tend to have better transferability and are less prone to overfitting. We believe that reducing the degree of fit of adversarial perturbations to specific models is beneficial to transferability.

\quad \noindent$\bullet$When models B and C exhibit high classification accuracy but low robustness, their decision boundaries are usually close. This means that the models make similar classification decisions on inputs. Under these circumstances, AEs can be more easily transferred from model B to model C because the perturbation direction in these examples typically follows the tangent direction along the decision boundary. We think that if the decision boundaries of models B and C are similar, the AEs will have equivalent effects on B and C.

\quad \noindent$\bullet$AEs can transfer between different models because they share a similar input space, i.e., they can process similar inputs. However, the transferability of AEs between models may be variable due to the wavy nature of decision boundaries. As decision boundaries change among models, the transferability of adversarial examples may be affected. Therefore, we consider it essential to understand the decision boundaries of different models and their relationship to the transferability of AEs.

\quad \noindent$\bullet$In deep learning models, the middle layers are critical for improving robustness and generalization. AEs and benign samples display noticeable differences in the middle layers of deep neural networks (DNNs), and these differences in middle-layer features are critical in determining whether an input is malicious or not.
We suspect that by harnessing the differences in middle-layer features, it may be possible to better distinguish between adversarial and benign samples, ultimately improving the robustness and transferability of AEs.

\section*{Transferability properties of audio AEs}
In this section, we illustrate the factors that affect the transferability of audio AEs and describe the geometric properties of the decision boundaries associated with speech recognition models.

\subsection*{\textbf{Potential factors impacting transferability}}
In our study, we pulled the code for attacking DeepSpeech\footnote{https://github.com/carlini/audio\_adversarial\_examples.} and Lingvo\footnote{https://github.com/cleverhans-lab/cleverhans/tree\\/master/cleverhans\_v3.1.0/examples/adversarial\_asr.} and replicated similar results as reported by the authors. Additionally, we contacted the authors of the ASpIRE model for attacking Kaldi \cite{35} and obtained some code and samples. Based on this, we conducted further research on adversarial attacks in speech recognition and observed the following characteristics:

1) We used the self-ensembling technique of dropout on the Kaldi ASpIRE model to produce AEs.  During the process of production, we queried the decoding results on the APIs every 10 iterations and found that after about 80 queries, the samples could be decoded as the target command on the Baidu Speech Recognition API \footnote{cloud.baidu.com}, Alibaba Cloud Speech Recognition API \footnote{cn.aliyun.com}, and Xfyun API \footnote{www.xfyun.cn}. The results are presented in Tab. \ref{Tab-noise}. This shows that a self-ensemble adversarial attack is a promising approach. We further investigated a more comprehensive ensemble adversarial attack that involves combining multiple base models to produce more transferable AEs.

The production of adversarial samples can be viewed as a training process, and there is just one input data set for training. With the dropout technique, during training, each neuron has its output set to 0 with probability $p$. In order to maintain consistency between the output during training and testing, it is usually necessary to divide the output by $1 - p$. But there is no testing involved in the training of adversarial samples, this step can be simplified to Eq.  \ref{EQ-dropout}:
\begin{equation}
\label{EQ-dropout}
\left\{ \begin{array}{l}
m \sim Bernouli(p)\\
\mathop x\limits^ \sim   = m \odot x
\end{array} \right.
\end{equation}
where $\odot$ denotes the multiplication operation and $\mathop x\limits^ \sim $ is the output; we do an Eq.  \ref{EQ-dropout} operation on the data before each iteration.

2) In adversarial attacks, adding noise is a common technique to increase the robustness of AEs, enabling them to successfully attack target models in a variety of settings. Several papers \cite{29,30,31,32} have demonstrated the effectiveness of noise in improving the robustness of models or AEs. Noise has a subtle effect, including preventing overfitting during training, producing highly robust, perturbation-resistant examples, and being used in both attack and defense. In experiments, we continuously increased the level of noise, adding random noise to data at each iteration during the production of adversarial examples, and then optimized according to the gradient. We found that the AEs performed significantly better as the noise levels rose. Although the number of iterations and time spent increased, the examples optimized with noise were easier to attack and more stable than those without noise. It became more difficult to produce adversarial examples with higher levels of noise, but once they were produced, the effectiveness of their adversarial attacks became much stronger.

\begin{table}[h!]
\caption{\label{Tab-noise} Results of transfer attack on API service.}
\begin{adjustbox}{width=0.48\textwidth}
\renewcommand{\arraystretch}{1.35}
\begin{tabular}{|c|c|c|c|}
\hline
\textbf{Noise}          & \textbf{Baidu-API} & \textbf{iFlytek-API} & \textbf{Alibaba-API} \\ \hline
\textbf{$\pm$0 }                   & 2/20      & 1/20      & 2/20       \\ \hline
\textbf{$\pm$4000}                 & 10/20     & 8/20      & 9/20       \\ \hline
\textbf{$\pm$8000}                 & 9/20      & 5/20      & 8/20       \\ \hline
\textbf{$\pm$12500  }              & 0/20      & 0/20      & 0/20       \\ \hline
\textbf{$\pm$17500}                & 4/20      & 2/20      & 5/20       \\ \hline
\textbf{$\pm$20000   }             & 12/20     & 12/20     & 11/20      \\ \hline
\textbf{$\pm$25000}                & 2/20      & 0/20      & 1/20       \\ \hline
\textbf{Dropout (p=0.5)}       & 13/20     & 12/20     & 11/20      \\ \hline
\textbf{Scale-Invariant (m=4)} & 14/20     & 13/20     & 12/20      \\ \hline
\end{tabular}
\end{adjustbox}
\begin{tablenotes}
\footnotesize
 \item[a]\textbf{Note:} The ``$\pm$N" indicates noise range is $[-N,N]$. The success rate of attack "A/B" indicates that there is A out of B AEs that trigger the command on the black-box platforms.
\end{tablenotes} 
\end{table}

In our experimental tests, we set the target command as "turn off the light" and used Eq. \ref{noise} to produce uniformly distributed random noise, repeating the experiment 20 times (with 20 input samples). This resulted in 20 adversarial samples. During the process of producing each AE, we queried the API every 10 iterations, totaling 200 queries. If the AE could be decoded as the target command by the API, we considered it to be an example of a successful transfer attack. As shown in Tab. \ref{Tab-noise}, without adding noise, the transferability of the AEs was very low, nearly zero. We also evaluated the perceptual quality of the samples, and there was almost no audible perturbation. As the noise level gradually increased to 20,000, the transferability gradually increased and then decreased, and the perceptual quality of the samples decreased significantly. When the noise level reached 20,000, the transferability was at its highest and the attack effect was at its best, but the noise completely covered the base carrier. As the noise level continued to increase, the attack success rate decreased sharply, and the transferability of the samples declined.
\begin{equation}
\label{noise}
f(x) = \left\{ \begin{array}{l}
\frac{1}{{2A}}, - A < x < A\\
0,else
\end{array} \right.
\end{equation}

Speech and image recognition systems exhibit differential sensitivities to contextual perturbations. While speech recognition is highly dependent on contextual cues, with minor fluctuations engendering significant decoding inaccuracies, image recognition is more robust to analogous variations. This disparity may stem from the non-linear relationship between noise level and transferability. Initial increases in noise may enhance the resilience of AEs, but continued escalation degrades contextual coherence to the point of outright obfuscation. Beyond a threshold noise level of 20,000, the contextual structure of speech signals is wholly disrupted, instead facilitating the emergence of adversarial contextual relationships. However, these maladaptive relationships also become increasingly difficult to establish as noise levels rise further.

3) Based on substantial testing, we found that the sensitivity of a model to different AEs is diverse, despite these having the same algorithm. The choice of audio carrier has a direct impact on the performance and transferability of the AEs, resulting in differential perturbation impacts and attack success rates. Some DL models appear inherently more sensitive to some input than others. For example, models tend to be more receptive to AEs crafted from images of snowy landscapes than from images of dogs. This effect is even more marked and extreme for speech recognition, with auditory perturbation and attack success rates showing significant differences due to the audio carrier. Conversational speech carriers tend to be more robust to attacks than music carriers, and simple musical elements are more readily perturbed than multi-element music carriers. To choose suitable carriers for generating AEs and to reduce the human perception of perturbations, research studies such as \cite{13,36}, and \cite{30} have used carrier selection methods such as automated selection and multi-angle evaluation.

4) Through our experimental testing, we found that the insertion of silence frames can improve the concealability of the target command and increase the attack success rate of AEs. Inserting silence frames in appropriate positions can significantly impact the decoding result of audio, which can be designed to enhance the robustness of adversarial attacks.

\begin{figure}[htbp]
\centering
\includegraphics[scale=0.5]{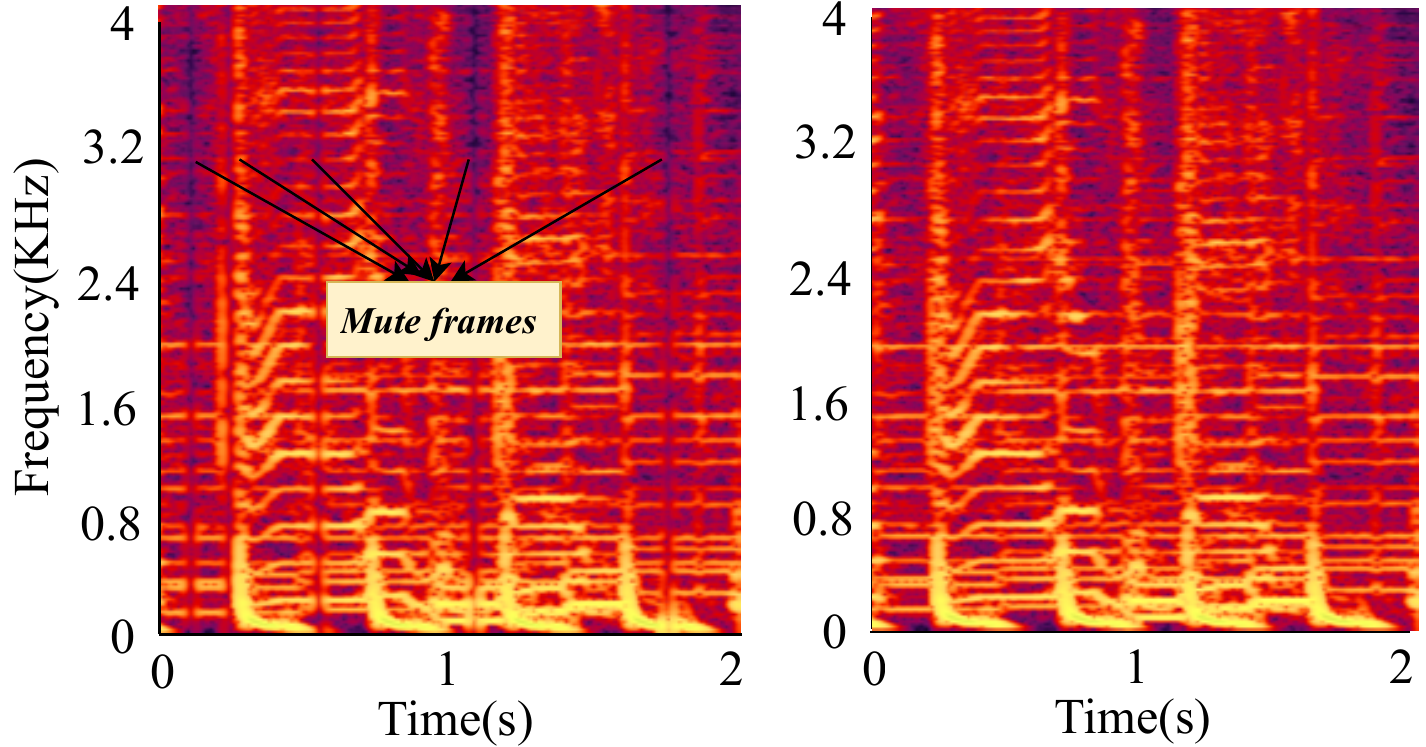}
\caption{\label{Noise}Compare the frequency domain features after the inclusion of silence frames.}
\end{figure}

In speech recognition, the input audio has to be segmented into frames. For a length of about 4 seconds of audio, it can usually be divided into about 500 frames. That means there are 501 possible insertion positions for an audio length of 4 seconds. At the same time, silence frames can disturb the decoded output of the whole window, and decouple the original contexts. The silence frames do not induce uncomfortable perception and only elicit minor pauses. We randomly picked 10 insertion positions during the perturbation overlay to study this effect. We found the effect to be more marked in the high-frequency domain (see Fig. \ref{Noise}) and weaker in the low-frequency domain for the inserted silence frames. The impact of carriers on AE is discussed in "CommanderSong  \cite{35}", carrier modality impacts AE feasibility, with some carriers enabling un notice AEs and others not by comparing their spectrums. We found that silence frame insertion is more akin to conversational carrier audio and could easily produce adversarial samples without being noticed. This may be because silence frames not only dissociate original contexts but also introduce blank segments, bringing the signal closer to conversational speech and increasing AE feasibility.

5) Scale-invariant adversarial attacks are a method of targeting models insensitive to scalar transformations. The scale-invariant adversarial attack is based on generating features that can fool models at various scales. \cite{31} generates scale-variant adversarial examples by processing samples at different scales, increasing transferability. \cite{32} proposes scale-invariant convolutional layers to enhance robustness by maintaining scale-invariant kernels, allowing recognition of similarly-featured inputs at different scales. Other work (e.g., \cite{22, 33}) applies scale and translation transformations to improve adversarial robustness at varying levels.
Recent work has shown that adversarial attacks that are invariant to scale transformations can be highly effective in damaging machine learning models. Scale-invariant attacks attempt to leverage the limited scale invariance of models by training adversarial samples across multiple scales. We study a specific scale-invariant adversarial attack approach with Eq. \ref{scale} and evaluate the effectiveness of speech recognition models. The results suggest that this scale-invariant approach achieves attack performance comparable to other adversarial attack methods. Furthermore, the produced adversarial samples demonstrate strong transferability to other models. For the optimal hyperparameter setting of m = 4, the attack achieves an average success rate of 65\% on deception transfer models. Overall, our results highlight the potential impact of scale-invariant adversarial attacks and the need for further research to alleviate such threats to model robustness.

\begin{equation}
\label{scale}
{g_t} = \frac{1}{m}\sum\limits_{i = 1}^m {{\nabla _x}l({S_i}({x^{adv}}),y)} 
\end{equation}
where  ${S_i}(x)$ denotes the scale copy of the input $x$  with the scale factor $1/{2^i}$  and $m$ denotes the number of the scale copies, $y$ is the true label.

\subsection*{\textbf{Geometric properties of ASR models}}
Briefly, the "transfer attack" means that the AEs produced in model A can successfully attack the targeted model (APIs). In our experiments, we analyzed how phonemes change from benign to audio AEs and found that some phonemes shift to the target phoneme with more difficulty than others, with varying patterns across models. We deduce that the decision boundaries between speech recognition models may be irregular and unstable, which is one of the reasons why speech recognition transferability attacks are difficult. 

\begin{figure}[htbp]
\centering
\includegraphics[scale=0.35]{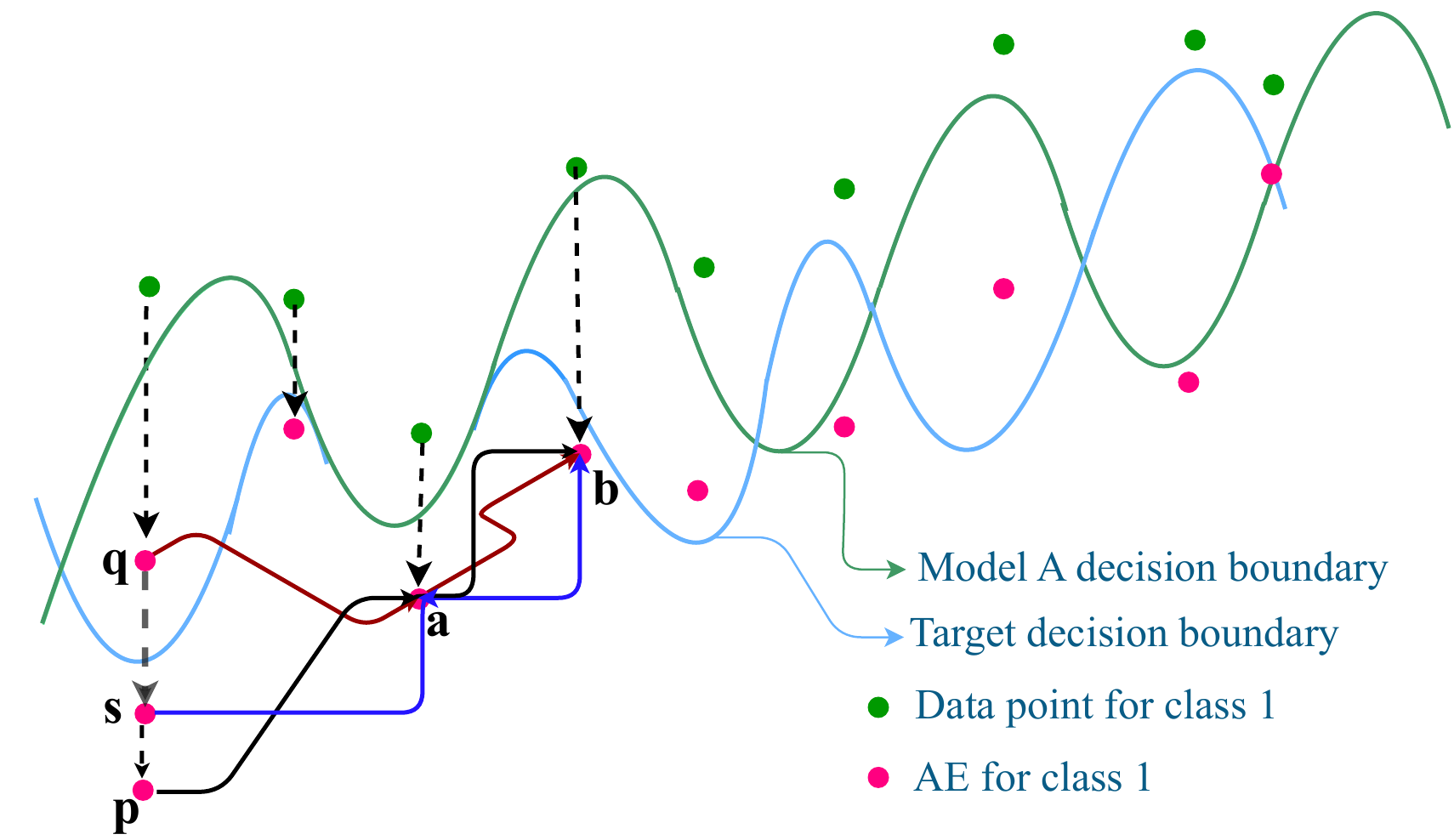}
\caption{\label{boundary}Transferable attacks on speech recognition models.}
\end{figure}

In Fig. \ref{boundary}, the green dots are data distribution points above model A's decision boundary, belonging to benign sample data points, and the input $x$ can be correctly transcribed as $y$. Adversarial samples (AEs, red points in the figure) formed from the benign sample points (green points) that have crossed the decision boundary. AEs are produced on model A, and the goal is to attack the target model. Although all green dots have crossed the decision boundary of A, some data dots (q) have not crossed the decision boundary of the target model and some have (s,p). For speech recognition, tri-phoneme correlation decoding is usually adopted, as shown in Fig. \ref{boundary}, where three data points with front and back correlations compose a decoding union. For a transferable attacker, to produce AEs that can attack the target on model A, they must cross the target's decision boundary for all data dots. Qab, pab, and sab are three decoding units, and pab, qab, and sab can be decoded into target commands on model A, but only pab and sab can be decoded into target commands on the target model. The main difference between pab and sab is the differing positions of p and s, suggesting that the level of perturbation is different. (The higher the perturbation, the lower the position.)

For example, in the words "end" and "old", the phoneme for the end is `IY-N-D', and for old is `UW-I-D.' Our goal is to transform "old" into "end." The essence of adversarial attacks in speech recognition is to make the phonemes cross the decision boundary of the target model to achieve the attack. The phonemes `I' and `D' correspond to the `a' and `b' dots in the figure, while `N' and `D' are the corresponding green dots above `a' and `b'. The dot q represents the phoneme "UW" for model A, but for the target model, only dots "s" and "p" are regarded as "UW." The qab cannot be decoded as "end" on the target model, and it may be decoded in other words. Even if the majority of phonemes ('N' and 'D') have crossed the decision boundary, it cannot ensure a target decoding result for "end" and "old." This is also the fundamental reason why transfer attacks are difficult.

In image or audio adversarial attacks, a "hard clip" is a technique used to limit the amount of perturbation within a certain range. Due to the measurement of auditory perturbation clips, in Fig. \ref{boundary}, the "s" and "p" point positions have different levels of perturbation, making it difficult to optimize to a specific point, such as the s-point, usually the p-point. Hard clipping is a rough way, and such clipped perturbation may result in uncomfortable listening, which is why many audio adversarial samples are noisy. Nevertheless, optimization to the s-point is always the desired goal. In addition, the choice of decoding rules will also impact the decoding results.

\section*{Enhance the transferability of AEs with the Ensemble-based approach}
Based on our findings, the dropout strategy incidentally increases the transferability of audio AEs. This encouraged us to further examine the impact of ensemble approaches on the transferability of audio AEs. We gathered Deepspeech, Kaldi, and Lingvo models as base models and evaluated the impact of different ensemble strategies on transferability.

\subsection*{\textbf{Rethinking the Ensemble learning}}
In supervised machine learning, we want to train a model that is stable and performs well in all aspects. In practice, this goal is often difficult to achieve, and we may only obtain several well-performing but still biased models. Ensemble learning is a technique that combines multiple weakly supervised models to create a more accurate and robust overall prediction. By combining them, the ensemble model can correct for the errors or biases in the individual models.

There are three approaches to ensemble learning: bagging, boosting, and stacking. Bagging uses random sampling with replacement to repeatedly extract different subsets from the training set to train a group of base models. Finally, the predictions of these baselines are combined using methods such as voting to reduce variance. Boosting trains the next base learner by focusing on the samples that were misclassified by the previous base model.  These results are weighted and averaged to produce the final model.  Stacking trains several base models using the training set and obtains the output results of these base models. These output results are then used as a new training set to train a model.

Similarly, ensemble learning has potential applications in attacking deep learning models, particularly black-box models. We can construct robust AEs by combining a collection of base models that can effectively attack black-box or gray-box models. These examples are then trained using a specific methodology to obtain embedded AEs. The gradient information collected from different models is used to generate an adversarial example based on a specified optimization method to attack the targeted model. It is expected that different ensemble methods will have varying impacts on gradient information.

\subsection*{\textbf{Ensemble attack}}
Different ensemble strategies have different impacts on the results of attacks. Below are some popular ensemble attack approaches.
\cite{3}  first implemented multi-model ensemble methods, for k base models, to obtain a loss ${J_i}(x)$ (after softmax) and a set of weights ${\alpha _i}$ to ensemble the optimization $\sum\limits_{i = 1}^k {{\alpha _i}{J_i}(x)} $. This approach soon became the baseline. Besides, various models have been used as base models for ensembling, while some have explored self-ensembling strategies. One example of such a strategy proposed by Yingwei Li et al. \cite{27} involves the use of ghost networks, which combine dropout, skip connection, and other operations to create a variety of candidate models that are then ensembled to produce AEs. In a related way, Dongxian Wu et al. \cite{28} proposed the skip gradient method, which employs a network with skip connections to give higher weight to shallow networks and attempt to produce AEs with stronger transferability. Another proposed Long-term Gradient Memory Ensemble Adversarial Attack \cite{23}, which is based on two assumptions: one that model transfer is equal to network generalization, and the other that boundary similarity is more important than perturbation limit s. SVRE-MI-FGSM \cite{24} also tries to improve the transferability of adversarial examples by reducing the variance between models, ultimately obtaining a smooth decision boundary curve to improve transferability.  The geometric properties of ImageNet were analyzed in \cite{25}, revealing that the gradient directions of different models are orthogonal to each other, and the decision boundaries of each model are relatively similar, facilitating the transferability of non-targeted attacks. Subsequently, two ensemble-based black-box attack strategies, the selective cascading ensemble strategy, and the stack parallel ensemble strategy were proposed in \cite{26} to implement more powerful black-box attacks on DL models. The diversity and number of substitutes in the ensemble are two important factors influencing the transferability of AEs, which is crucial for selecting effective substitutes for the ensemble.

\begin{figure}[htbp]
\centering
\includegraphics[scale=0.35]{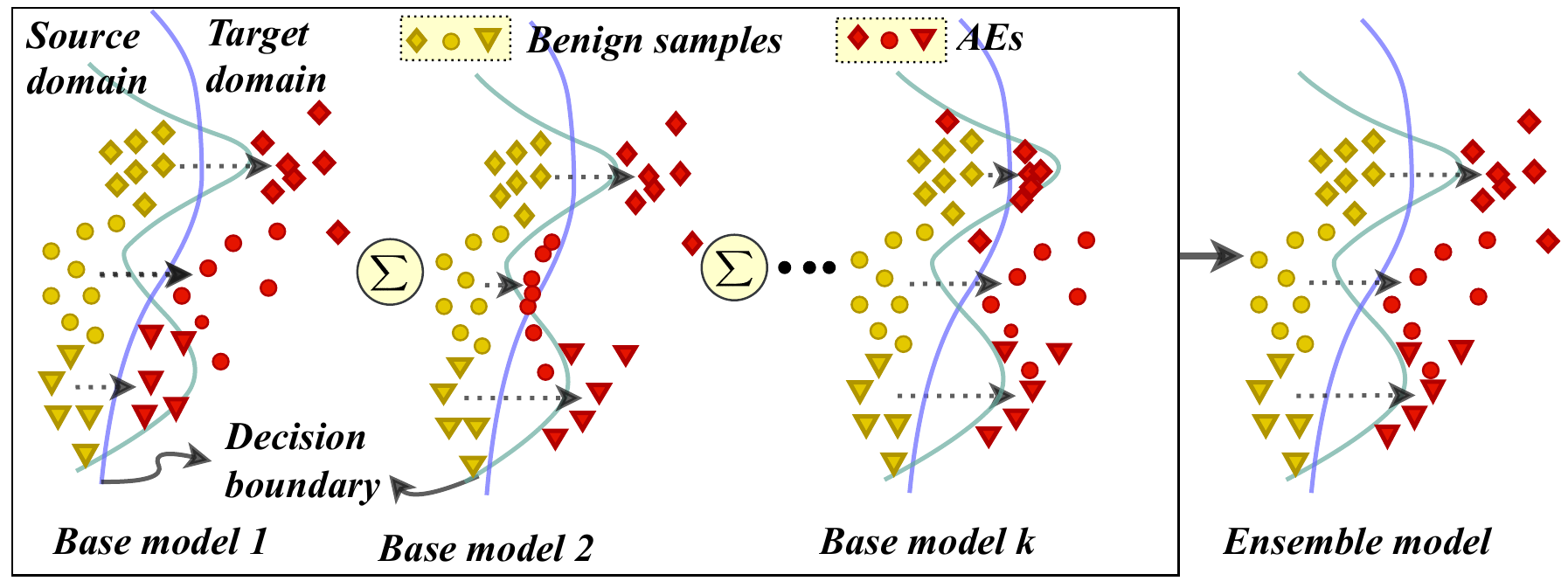}
\caption{\label{decision}Transferable attacks on speech recognition models. Yellow icons mark benign samples and red icons mark adversarial samples.}
\end{figure}

For the adversarial attack on a single model, it is only necessary to cross the decision boundary of that specific model. However, there is a degree of gap between the decision boundaries of different models. As illustrated above, the decision boundary in the speech recognition model is sharply wavy and irregular. The ability of different models to cross the classification boundary is not equal, and ensemble methods can help to overcome this deficiency. As shown in Fig. \ref{decision}, AEs produced based on a single model 1 may not transfer successfully to other models due to differences in decision boundaries. The same problem is often present in the production of AEs based on other models. If those basic models are connected, it is possible to obtain AEs that cross more model decision boundaries, thus forming stronger attack capabilities.

It can be observed that the AEs (the red dot) generated by the ensemble-based strategy cross $k$ decision boundaries of substitute models, indicating its higher generalization ability and a greater likelihood of successfully deceiving the target black-box model API, i.e., crossing the decision boundary of the target model.

\subsection*{\textbf{Ensemble-based black-box API attack strategies}}

\subsubsection{Models and Attack Approach}
As previously proposed, self-ensemble dropout technology has somehow improved the transferability of AEs. Despite the irregular and unstable decision boundaries that exist between the source model and target model, the transfer of AEs has proven to be difficult. However, the ensemble method can continue to close toward the best overlapping subspaces and the "pab" attack pathway. Nonetheless, it is important to note that self-ensemble dropout technology may close the "sab" attack pathway, which can marginally impair auditory perception.
Adversarial attacks on speech recognition models are more complicated than those on images, and the techniques probably vary considerably. However, no one has attempted to generate AEs by combining different speech recognition models. This is partly due to the scarcity of open-source speech recognition models, as well as the greater complexity of these models, which require more time and resources to produce AEs. Also, the non-smoothness of decision boundaries in speech recognition models poses a challenge to achieving the desired transferability. To further investigate the impact of the ensemble on AE transfer, we examined the Deepspeech, Kaldi, and Lingvo models and designed a combined algorithm, with experimental details explained in the following section. The Deepspeech, Kaldi, and Lingvo models are currently the most popular neural network-based speech recognition models and have been widely accepted by both industry and academia.

In our study, we selected these three models as base models for ensemble research. 

To attack the Deepspeech model, we used the CTC-loss of the Deepspeech model as the loss function and optimized it through Adam. The method for obtaining AEs is shown in Eq. \ref{dp}.

\begin{equation}
\label{dp}
x_t^{adv} = x_{t - 1}^{adv} + clip_x^\varepsilon (Adam({\nabla _x}los{s_{ctc}}(x_{t - 1}^{adv},y)))
\end{equation}

AEs can be generated to attack Kaldi's Aspire model by computing the pdf-id loss function between the input sample $x$ and the target $y$ and performing gradient descent with the Adam optimizer. The maximum value of the perturbation is also constrained. The implementation is shown in the following Eq. \ref{kd}.

\begin{equation}
\label{kd}
x_t^{adv} = x_{t - 1}^{adv} + clip_x^\varepsilon (Adam({\nabla _x}los{s_{{\rm{pdf}} - ids}}(x_{t - 1}^{adv},y)))
\end{equation}

The adversarial attack on the Lingvo model is a two-step attack algorithm. In the first stage, produce a AE that can fool the Lingvo model by using the Adam optimizer and gradient descent to according the gradient. In the second stage, the AEs are continuously optimized using the psychoacoustical auditory masking principle. Adversarial perturbations are injected in ranges that humans cannot perceive. This principle is shown in Eq. \ref{lg}.

\begin{equation}
\label{lg}
\left\{ \begin{array}{l}
x_{1t}^{adv} = x_{1(t - 1)}^{adv} + clip_x^\varepsilon (Adam({\nabla _x}los{s_{\bmod el}}(x_{1(t - 1)}^{adv},y)))\\
x_{2t}^{adv} = x_{1t}^{adv} + Adam({\nabla _x}los{s_{masking}}(x_{1t}^{adv},{y_{masking}}))
\end{array} \right.
\end{equation}

\subsubsection{ Ensemble Attack Methods}
 
In the image domains, for each model, the ensemble attack employs the same attack method (the FGSM method is commonly adopted). On this basis, there are three different levels of the ensemble, one in predictions, the other in logits, and the third in loss functions.
The methods are shown in Eq. \ref{ens}.

\begin{equation}
\label{ens}
\begin{array}{l}
l(x_t^{adv},y) =  - {1_y}*\log (\sum\nolimits_{k = 1}^n {{\alpha _k}{p_k}(} x_t^{adv}))\\
l(x_t^{adv},y) =  - {1_y}*\log (soft\max (\sum\nolimits_{k = 1}^n {{\alpha _k}\log it{s_k}(} x_t^{adv})))\\
l(x_t^{adv},y) = \sum\nolimits_{k = 1}^n {{\alpha _k}{l_k}(} x_t^{adv},y))
\end{array}
\end{equation}
where  $l$ is the loss of the model, ${1_y}$  is the one-hot encoding of the ground-truth label $y$ of $x$,  ${p_k}$ is the prediction of the k-th model, and ${\alpha _k} \ge 0$ is the ensemble weight constrained by $\sum\nolimits_{k = 1}^n {{\alpha _k} = 1} $.

\begin{figure}[htbp]
\centering
\includegraphics[scale=0.35]{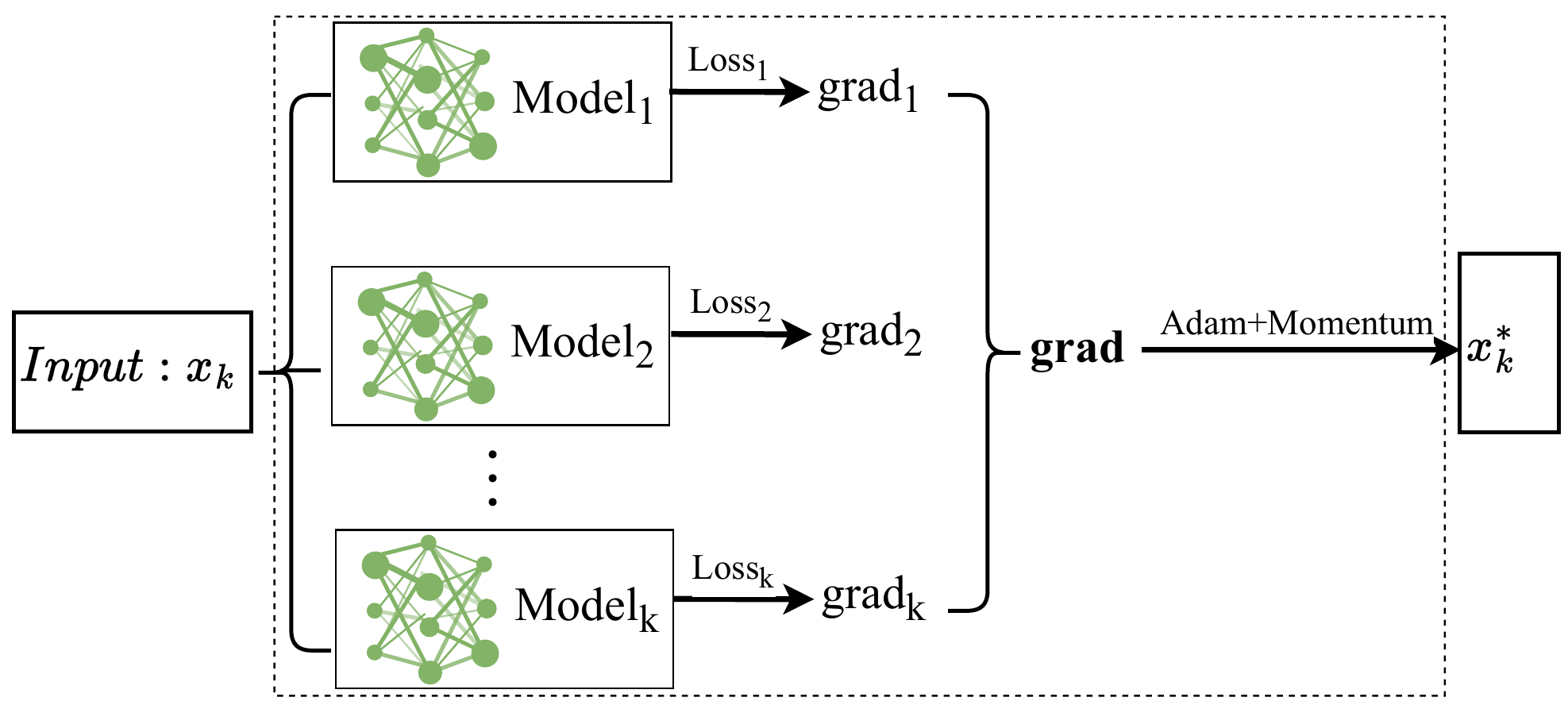}
\caption{\label{strategy}Adversarial sample generations based on ensemble attacks.}
\end{figure}

In audio adversarial attacks, a host of methods has been developed to target particular models, frequently relying on distinct loss functions to produce different attack strategies and outcomes. However, the mainstay of these attacks relies on the use of gradient descent, which means that they all need to calculate gradients and thus pose an opportunity for gradient-based embedding. In this study, we investigate two gradient ensemble strategies, the serial and parallel gradient ensembles, to produce highly transferable AEs. Specifically, we employ two ensemble algorithms to compute different gradients for a given input based on the base model and combine them according to the ensemble strategy to update AEs with gradient momentum and the Adam algorithm. The conceptual framework is shown in Fig. \ref{strategy}.

Moreover, the transferability of AEs in speech recognition is significantly inferior to images due to the more oscillatory nature of the decision boundary. This irregular and unstable boundary hinders the transferability of AEs. However, local smoothing has the advantage of reducing the violent oscillations of the decision boundary. In \cite{42} and \cite{43}, the authors propose adding a slight degree of noise to the input data and repeatedly applying it to smooth the impact of noise in models. Alternatively, \cite{44} smooths the input image and use a gradient ascent algorithm to maximize the error rate of the predictive classifier, resulting in a more valid and highly transferable attack on the classifier.

In the 2017 NIPS competition, Dong et al. \cite{41} demonstrated that gradient momentum significantly improves the transferability of samples, enabling better avoidance of local minima. Specifically, the gradient momentum algorithm adds the current gradient vector to the previous gradient vector and multiplies it by a momentum factor before updating the parameters at each iteration. This approach maintains earlier gradient information, allowing the model to maintain a certain direction and speed during descent, thereby increasing the convergence rate.

In our study, we propose a novel gradient smoothing method to enhance the adversarial transferability of ensemble attacks and address the issue above. Our approach is inspired by the Gaussian noise gradient smoothing method \cite{42,43,44}, which designs the gradient by taking into account the gradients of other points around the modified point and treats the averaged value of the gradient after multiple Gaussian fuzzes as the final gradient. Our methodology involves multimodel gradient smoothing for each iteration. First, superimpose white noise on the input, the sound is susceptible to disturbance from white noise rather than Gaussian noise. Next,  sum the gradient information for each model, respectively, and finally, normalize the summed gradient to obtain the final smooth gradient. To increase the number of models, we combine the self-ensemble dropout method to enhance model diversity. The calculation process for our practice is presented in Eq. \ref{g} and Eq. \ref{eq_norm}.

\begin{equation}
\label{g}
\begin{array}{l}
{{\tilde g}_t} = \sum\limits_{i = 1}^m {{\nabla _x}l(D(x_t^{adv})}  + {\varsigma _i},y),{\rm{  }}{\varsigma _i} \sim U( - A,A)\\
\end{array}
\end{equation}
\begin{equation}
\label{eq_norm}
{g_t} = {{\tilde g}_t}/means(abs({{\tilde g}_t}))
\end{equation}
where $D$ is the dropout function. 

After the smoothed gradient ${g_t}$ is obtained, we employ Eq. \ref{m} to accelerate the convergence speed and overcome the issue of slow and unstable convergence during gradient descent, thereby promoting a smoother optimization process and generating high-quality AEs more quickly.

\begin{equation}
\label{m}
{g_{t + 1}} = \mu {g_{t - 1}} + {g_t}
\end{equation}

\subsubsection{Ensemble strategy}
1) \textbf{Random Gradient Ensemble (RGE)}. For gradient ensembles, there are serial and parallel ensembles. The serial ensemble adversarial attack relies on diverse base models, randomly picking the gradient of one model as the ensemble gradient for each iteration, and after T iterations, getting terminal AEs. Ensuring model diversity is a crucial factor in ensemble attacks, and a random gradient can increase model diversity. Randomly selecting a model's gradient can increase uncertainty and escape over-fitting to a specific model, thus improving transferability. Moreover, different models may have different levels of robustness, and by randomly selecting a model's gradient, it is possible to expand between different models and identify robust models, which may help to improve the robustness of AEs. The random gradient ensemble approach has also been implemented in \cite{61} for ensemble adversarial training, and has achieved positive results. The algorithmic workflow is outlined in algorithm \ref{alg}.

2) \textbf{Dynamic Gradient Weighting Ensemble (DGWE)}. In general, the larger the gradient value (L-norm), the more important its role in updating model parameters is, and it dominates the process. Yet, in ensemble training, when the model's gradient value and direction differ substantially from those of other models, the ensemble model is more susceptible to being steered by large-gradient models, causing the model's predictions to lean towards such models and failing to achieve the desired ensemble. This issue may arise because large gradients contain noise or occur in irregular regions, which have a heavier impact on parameter updates. Larger gradients may also indicate that the input data is less reliable, whereas smaller gradients imply greater reliability. The same challenge arises in ensemble adversarial attacks, where assigning lower weights to models with gradients that differ from those of other models may prevent their overwhelming influence on the final output, hence promoting the diversity and robustness of AEs.
In our DeepSpeech, Kaldi, and Lingvo repositories, we measured considerable variation in gradients among the three models. Consequently, it is important to avoid over-reliance on individual models in the ensemble. We recommend assigning lower weights to models with larger gradients to limit their undue influence on the ensemble. We found that this approach can also reduce the variance of the ensemble, ultimately leading to lower gradients for parameter updating and avoiding the potential for destructive bias from larger gradients.

 Here, we propose DGWE based on the $|| * |{|_2}$ norm of the gradient. The $|| * |{|_2}$ norm of the gradient provides a weighting formula for input, depending on its importance to the result. The specific setting method is as follows Eq. \ref{weight}:

\begin{equation}
\label{weight}
\left\{ \begin{array}{l}
{w_i} = \exp ( - ||{g_{tk}}|{|_2}^{1/{\sigma ^2}})\\
{w_i} = {w_i}/\sum\limits_{i = 1}^K {{w_i}} 
\end{array} \right.
\end{equation}
in this method, $||{g_{tk}}|{|_2}$ represents the gradient Euclidean norm of model $k$, while $\sigma$ is a parameter used to control the smoothness of the weights (in this paper, we take  $\sigma $=1). When $\sigma$ is large, the weights become smoother, resulting in smaller differences in weights among different data sources. Conversely, when  $\sigma$ is small, the weights become more dynamic, resulting in larger differences in weights among different data sources. This method is an advanced data fusion technique that dynamically adjusts the weights based on the contribution of each model's gradient to the outcome, leading to a more refined overall gradient that is better suited for the optimization process. The algorithm flow is described in the DGWE part of algorithm \ref{alg}.

\begin{algorithm}[!tb]
\SetAlgoLined
\caption{\label{alg}The Audio Ensemble attack algorithm} 
\LinesNumbered
\KwIn{A benign audio example $x$, and its label $y$, a set of K base models and the corresponding gradients $\left\{ {{g_1},{g_2}...{g_k}} \right\}$, an ensemble gradients ${g^{ens}}$. The perturbation bound $\varepsilon$, number of iterations T, internal update frequency M.}
\renewcommand{\baselinestretch}{1.5}
\KwOut{An adversarial example  ${x^{adv}}$, that fulfills $|{x^{adv}} - x|{|_2} \le \varepsilon $}.
Initialize $x_0^{adv} = x$, 

\For{$t = 0$ to $T - 1$}{
Pick model index $k \in \{ 1,2...K\}$

Gradient Smoothing,

\For{$m = 0$ to $M - 1$}{
Initialize ${g_{0k}} = 0$

Calculate the gradient of the base model:

$\begin{array}{l}
{{\tilde g}_{mk}} = {\nabla _x}l(D(x_t^{adv}) + {\varsigma _i},y),{\rm{  }}{\varsigma _i} \sim U( - A,A)\\
{g_{mk}} = {g_{(m - 1)k}} + {{\tilde g}_{mk}}
\end{array}$
}
Normalization,

${g_{tk}} = {g_{mk}}/means(abs({g_{mk}}))$,

Reset  ${g_{mk}} \leftarrow 0;$

Momentum gradient,

${g_{tk}} = \mu {g_{(t - 1)k}} + {g_{tk}}$.

\# The method of RGE.

Produce a random index ${\rm{r}} \in \{ 1,2...K\}$,

Make ${g^{ens}}{\rm{ = }}{g_{rk}}$.

\# The method of DGWE.

Get the weight   by Eq. \ref{weight}.

${g^{ens}} = \sum\limits_{k = 1}^K {{w_k}{g_{tk}}}$.

\# Update the outer adversarial example,

$x_{t + 1}^{adv} = x_t^{adv} + clip_x^\varepsilon (Adam({g^{ens}}))$.
}
 \Return ${x^{adv}} = x_T^{adv}$.
\end{algorithm}

\section*{Experiments and evaluation of ensemble attacks}

\subsection*{\textbf{Evaluation Metrics}}
A transferable adversarial attack is an approach that encompasses reduced expenditure of time and resources for generating AEs as well as potential enhancement of attack success rates. However, it is essential to assess the transferability of the attack, as the ability of a transferable adversarial attack could be negated by shifts in the target model. In order to evaluate the transferability of the AEs generated by the RGE and DGWE approaches, we measured the transfer rate (TR) of the AEs \cite{46,47}.TR serves as a metric that defines the misclassification rate of AEs generated by the local model relative to the target model. Superior transferability is indicated by a higher transfer rate.

In our experiments, we trained 20 AEs on the ensemble model (single input, requiring 20 manual runs of the experiment), with every 5 samples corresponding to one target command (4 target commands in total). In this work, we do not consider the influence of target commands on transferability \cite{48}, despite the possibility that different target commands have different transferability. Our primary goal is to evaluate the transferability of the AEs generated by the ensemble method, emphasizing overall transferability. We consider the ratio of the number of AEs containing different target commands that are successfully transferred to the target model divided by the total number of AEs as the transfer rate to assess the transferability performance of our method, as shown in Eq. \ref{TR}.

\begin{equation}
\label{TR}
TR = \frac{{\sum\limits_{i \in T}^{} {\sum\limits_{j \in N}^{} {F({x_{ij}})} } }}{X}
\end{equation}
where $T$ is the target command group, $N$ is the sample group of a specific command, ${x_{ij}}$ indicates the j-th sample of the i-th command, $X$ is the total sample count, and $F(.)$ is the target model output. In Tab.  \ref{Tab-result}, we usually display this TR value in "A/B" form (which can also be converted to decimals).

\subsection*{\textbf{Experiment Setting}}
To evaluate the transferability of our proposed combined adversarial attack strategy, we conducted experiments on commercial speech recognition APIs, such as those provided by iFlytek, Alibaba, and Baidu. These APIs provide high-level English speech recognition services that directly impact the user experience of millions of people. Attacks against these commercial APIs are closer to real-world attack patterns and are more sophisticated. In the choice of attack targets, we picked some representative commands as attack targets, \footnote{including turn off the light, "open a website, where is my car, and what is the weather}. In this context, higher transferability implies a greater potential for danger.

Also, as shown in the previous experiments, the carrier of AEs plays a significant role in transferability. The generation and transferability of AEs are directly influenced by the choice of carrier. In CommanderSong \cite{35}, Cheng Yuxuan et al. first used music segments as the carrier of AEs in attacks, arguing that music has the nature of common consumption, giving it native opportunities in attacks with its popularity and extensive reach. Attacks on music segments are likely to raise public concern. Music segments have also been shown to serve as carriers in \cite{39,49}. In our study, we also consider music segments as carriers of AEs, including popular music, classical music, rock music, and light music, covering multiple language types such as Korean, English, Japanese, Chinese, Russian, and Arabic. The length of each music segment is about 4 seconds.

\subsection*{\textbf{Evaluation of the attack}}
In this section, we conduct a quantitative analysis of the transferability differences between RGE and DWGE by comparing several state-of-the-art attacks, including DS \cite{34}, ITRA \cite{15}, and CS \cite{35} for white-box attacks, and DW \cite{39}, Occam \cite{40} and DC \cite{49} for black-box attacks on commercial APIs. This comparison provides additional support for investigating the interpretability and robustness of the models. All of these attacks are single-model transfer attacks in speech recognition adversarial attacks and do not exploit ensembles. This is possibly overly expensive, and there is also a lack of available models that can be combined. There are still many gaps in exploring the transferability of adversarial examples trained from ensembles. In this study, we strive to meet these gaps by gathering three speech recognition models capable of supporting ensemble attacks. We also propose an attack algorithm for speech recognition model ensembles. Tab. \ref{Tab-result} shows the transferability of each attack to different APIs.

\textbf{DS attack}.The Deepspeech Attack (DS) \cite{34} is a white-box attack against the DeepSpeech model. There are two methods to train adversarial audio, namely gradient descent and genetic algorithms. Both methods are time-consuming. The DS attack can generate an audio waveform that is 99.9\% similar to the original, which can be transcribed into any target phrase. The DS attack was first implemented in DeepSpeech-0.1.0 and has been found to exhibit good transferability to versions 0.2.0 and 0.3.0 \cite{48} This can be attributed to the similarity in model structure and optimization algorithms in these versions. However, the transferability of the DS attack to APIs is poor, with an average transfer rate of only 0.1. This limitation makes it difficult to extend the attack to other models.

\begin{table*}[tbp]
\centering
\caption{\label{Tab-result}Transfer Rate (TR) of Adversarial Samples on Commercial Cloud Speech-to-Text APIs.}
\begin{adjustbox}{width=0.95\textwidth}
\renewcommand{\arraystretch}{1.25}
\begin{tabular}{|c|c|c|c|c|c|c|c|c|c|}
\hline
\textbf{API service}   & \multicolumn{1}{l|}{\textbf{Audio length}} & \textbf{DS} & \textbf{ITRA} & \textbf{CS} & \textbf{DW} & \textbf{Occam} & \textbf{DC} & \textbf{RGE} & \textbf{DGWE} \\ \hline
\textbf{Aliyun-API}    & 4s                                         & 1/20        & 0/20          & 4/20        & 6/20        & 20/20          & -           & 14/20          & 8/20           \\ \hline
\textbf{Xfyun-API}     & 4s                                         & 0/20        & 0/20          & 8/20        & 14/20       & 20/20          & 13.8/20     & 12/20          & 14/20          \\ \hline
\textbf{Baiduyun-API} & 4s                                         & 2/10        & 0/20          & 2/20        & -           & -              & 14.9/20     & 15/20          & 13/20          \\ \hline
\textbf{Tencent-API}   & 4s                                         & 1/20        & 1/20          & 2/20        & 8/20        & -              & 16.5/20     & -              &                \\ \hline
\textbf{Microsoft-API} & 4s                                         & 6/20        & 3/20          & 6/20        & 16/20       & 20/20          & -           & -              &                \\ \hline
\textbf{Average}       & 4s                                         & 2/20        & 0.8/20        & 4.4/20      & 11/20       & 20/20          & 15.1/20     & 13.7/20        & 11.7/20        \\ \hline
\end{tabular}
\end{adjustbox}
\begin{tablenotes}
\footnotesize
 \item[a]  \textbf{Note}: The abbreviations DS, ITRA, CS, and DW refer to different attack methods presented in \cite{34}, \cite{15}, \cite{35}, and \cite{39}, respectively. Occam and DC refer to the attack methods proposed in \cite{40} and \cite{49}, respectively. The "-" indicates that the attack method has not been tested on the API. In the DW \cite{39}, the authors evaluated 10 AEs. To match the number format of this work, we doubled the number. In the DC \cite{49}, the authors reported the attack success rate as the result, which was converted by calculation to the format.
\end{tablenotes} 
\end{table*}

 \textbf{IRTA attack}.The IRTA attack algorithm \cite{15} is a two-stage method designed to target the Lingvo white-box model. In the initial stage, a sample is generated to deceive the Lingvo model. In the subsequent stage, psychoacoustic masking is applied to inject adversarial perturbations in the audio range that are imperceptible to humans. Compared to other approaches, IRTA is remarkably quiet, and the perturbations are barely detectable to the human ear. This feature is closely related to the human hearing range. Regrettably, our investigation revealed that the transferability of IRTA's AEs is seriously limited, with an average transferability rate of only 0.04. This implies that this method is only effective for specific models. The primary reason for this limitation is that the automatic speech recognition (ASR) model focuses solely on the human audible range and is indifferent to the inaudible range. Consequently, sound in this range is also ignored by the ASR model.

\textbf{CS attack}. The CommanderSong Attack (CS) \cite{35} is a white-box attack that embeds a command into a song. This manipulated song can then be played to activate real-world ASR systems and execute targeted commands. In the context of Kaldi, the WTA attack can achieve a 100\% success rate, while the WAA attack can achieve a 96\% success rate. Our investigation revealed that the average transfer rate (TR = 0.22) of CS adversarial examples (AEs) is higher than that of DS and ITRA, potentially indicating a higher degree of perturbation. Through an analysis of AEs and codes, we observed that the degree of clip perturbation in the CS attack is greater than that of DS and ITRA (approximately 4000 for CS compared to 2000 for DS and ITRA). So, the auditory perceptibility of the CS attack is inferior compared to the DS and ITRA.

\textbf{DW attack}. The Devil's Whisper (DW) attack \cite{39} is designed to target commercial black-box models that deploy surrogate models to estimate the target model. The AEs can successfully attack various commercial devices, including Google Assistant, Google Home, Amazon Echo, and Microsoft Cortana, with an average success rate of executing 98\% of target commands. However, the transfer performance of DW is relatively high (TR = 0.55) for iFlytek, while poor for Tencent and Alibaba APIs. DW is an attack on APIs that poses a low risk of model overfitting. The training of AEs can be tuned based on feedback from different APIs to attack them more effectively, achieving a higher transferability rate than other attacks. However, due to differences in models, generating AEs based on the feedback of a specific API may lead to stronger attack capabilities against some APIs and weaker against others.

\textbf{DC attack}. The Disappeared Command (DC) attack \cite{49} relies on an acoustic masking model to attack ASR systems. By injecting noise or inserting additional audio to conceal the target command and prevent it from being recognized as normal speech. In their study, the authors evaluated the effectiveness of AEs generated with the DC attack on iFLYTEK, Tencent, and Baidu APIs and achieved promising results. However, the author did not release the source code or samples, and we have not verified their accuracy or conducted further evaluations on other API platforms.

\textbf{Occam attack}. The Occam attack \cite{40} generates AEs through a coevolving optimization algorithm, which enables the identification of optimal solutions with minimal information and effectively fools speech recognition systems. While the success rate of the attack on APIs can reach 100\%, the authors have not released the source code or provided access to the AEs, limiting further evaluation and confirmation of the reported results.

\textbf{The evaluation of RGE and DGWE attack}. The ensemble attack with RGE is a serial method that randomly selects gradients as the ensemble model gradient to update ${x^*}$. As shown in Table \ref{Tab-result}, the ensemble adversarial method of RGE exhibits remarkable transferability to the Baidu, Alibaba, and iFlytek APIs, with an average transfer rate of 0.69, which surpasses that of DS, ITRA, CS, and DW, but falls short of Occam and DC. This finding highlights the potential of ensemble attacks in enhancing the robustness of samples and promoting generalization to other models. Besides, our analysis suggests that the similarity of decision boundaries in speech recognition models has a limited impact on the transfer rate of samples. As discussed in \cite{48}, the robustness of sample features may play a more crucial role in transferability than model similarity, and randomization can mitigate the risk of overfitting and improve the robustness of AEs. Overall, the results show that ensemble attacks with random gradient ensembles can effectively improve the transferability of AEs to normal black-box APIs.

In Table \ref{Tab-result}, the transfer rate of the parallel ensemble attack method based on DGWE is inferior to that of the serial ensemble method based on RGE, with an average transfer rate of 0.59, which is about 0.1 lower than RGE. Notably, both RGE and DGWE present poorer performance on Alibaba's API compared to iFLYTEK and Baidu, suggesting that different APIs have different susceptibilities to AEs.
The experiments suggest that black-box models with high accuracy are highly susceptible to the effects of RGE and DGWE proposed in our study, and any ensemble method can have a positive impact on the transferability of AEs. We also observed that carefully crafted AEs with high success rates possess stronger transferability. RGE is a more effective method than DGWE and can generate AEs with stronger transferability.

\subsection*{\textbf{ Transferability factors exploration}}
To address the lack of diversity in open-source speech recognition models based on deep learning, we propose a method that combines dropout self-ensemble and model ensembling to increase model diversity. Specifically, the dropout randomly sets partial inputs to zero with probability $p$, triggering the death of some neurons during propagation and consequently the formation of multiple networks. In adversarial attacks, the value of input $x$ is updated, serving as both an input and a parameter update. We randomly set a portion of $x_i$ to zero with probability $p$, and subsequently update $x$ via the gradient of the new input, as outlined by Eq. \ref{eq-dropout}.

\begin{equation}
\label{eq-dropout}
\left\{ \begin{array}{l}
{r_i} \sim Bernoulli(p)\\
x_0^{adv} = x\\
x_t^{adv} = x_{t - 1}^{adv} + clip_x^\varepsilon (Adam({\nabla _x}los{s_{ctc}}({r_i}*x_{t - 1}^{adv},y)))
\end{array} \right.
\end{equation}
where ${r_i}$ represents a Bernoulli distribution with the same dimension as $x$, and the probability parameter $p$ plays a crucial role in the dropout function. If $p$ is set too low, the number of models in the ensemble may be insufficient, thereby impairing the effectiveness of the ensemble. Conversely, if the value of $p$ is excessively high, the sample training process may diverge, resulting in non-convergence or high convergence costs, ultimately leading to the training of samples with weak transferability.

\begin{figure}[htbp]
\centering
\includegraphics[scale=0.35]{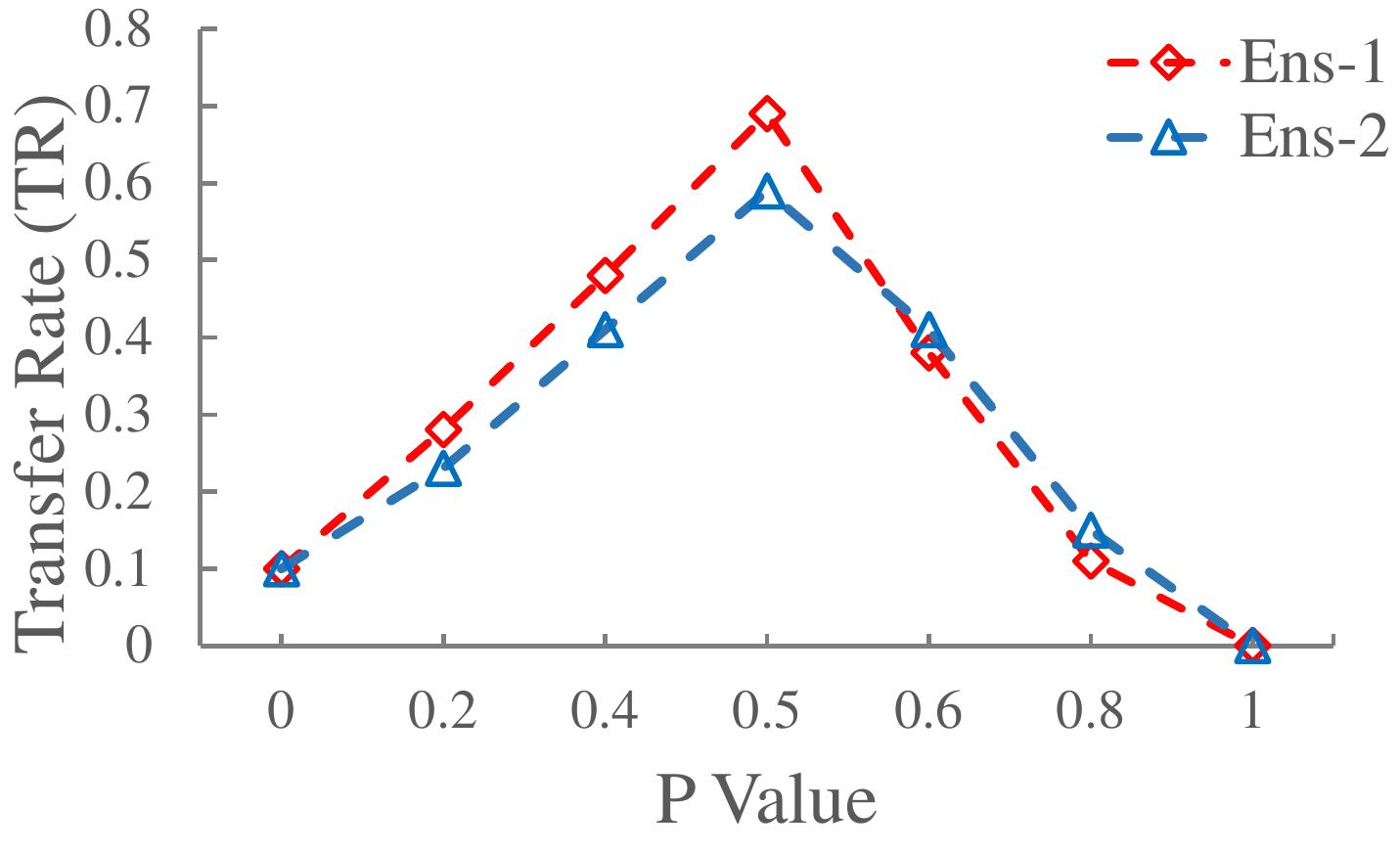}
\caption{\label{fig-dropot}The transfer rate (TR) as a function of the dropout rate ($p$ value) and shows an increasing trend with rising $p$ value until it culminates at a $p$ value of 0.5. Subsequently, the TR experiences a decline with the $p$ value continues to increase.}
\end{figure}

The experimental results of the relationship between $p$ value and transferability are shown in Fig. \ref{fig-dropot}. As the $p$ value increases, the transferability of both ensemble methods shows a tendency to increase and then decrease. The transferability is up to a peak at $p = 0.5$, where RGE achieves a transferability of 0.69 and DGWE 0.59, with a difference of only 0.1. Initially, the transferability is positively correlated with the $p$ value, since a higher $p$ value implies an increased number of ensemble models, thus slightly decreasing the degree of fitting of the sample to a particular model, and improving the robustness of the AEs. However, at a certain point, the transferability starts to decrease with the $p$ value. When $p=1$, ${r_i}$ is 0, the input becomes 0 and the parameters are not updated, resulting in a transferability of 0. In the process of decreasing transferability, RGE decreases faster than DGWE and becomes 0 earlier. The continuous increase in the value of $p$ makes the loss function harder to converge and raises the cost of obtaining samples, leading to poorer results.

\section*{Discussion}

\subsection*{\textbf{Defense}}

The ASR system is a sophisticated model that consists of several components, including preprocessing, feature extraction, acoustic processing, and language processing, each of which is vulnerable to attacks, as is the defense. Adversarial training, input preprocessing, model adaptation, detection, and randomization are commonly implemented as defense mechanisms.

Previous studies \cite{17} and \cite{18} suggest training an ASR with adversarial audio can help defend against certain attacks. However, the cost of creating such adversarial audio is expensive. Additionally, this approach can only resist attacks that have already been identified; it is not capable of preventing new attacks. Furthermore, since most adversarial audio is generated from music clips, it is difficult to evaluate the effectiveness of adversarial training on ASR models. Therefore, the evaluation of adversarial training poses a significant challenge.

In the literature, several defenses have been proposed to deal with adversarial perturbations in ASR systems. Randomized smoothing is proposed in \cite{10}, while \cite{11} suggests the WaveGAN vocoder reconstructs the waveform and removes the perturbation. Other methods, such as label smoothing \cite{12} and audio compression \cite{13,14} are also recommended. Additionally, downsampling methods \cite{15} and appending distortion signals \cite{16} have been explored.

In this study, we evaluate the effectiveness of two defense methods, downsampling defense \cite{39} and noise defense \cite{35}, In the context of three ASR APIs (Baidu, Alibaba, and iFlytek),  we tested the downsampling resampling rate between 5200 Hz, 5600 Hz, and 6000 Hz, and the noise amplitude between 500, 1000, and 2000 based on the 20 AEs. The results are presented in Table \ref{Tab-defense}, which denotes that the downsampling defense outperforms the noise defense. The effectiveness of the downsampling method can be attributed to its ability to compress the data and interpolate for reconstruction, making it difficult to restore the carefully injected perturbation and filter high-frequency signals. On the other hand, the noise defense method masks the adversarial perturbation with random noise, but this approach can also distort auditory perception.

\begin{table}[!h]
\caption{\label{Tab-defense}Results of defense.}
\begin{adjustbox}{width=0.48\textwidth}
\renewcommand{\arraystretch}{1.35}
\begin{tabular}{|c|c|c|c|c|}
\hline
\textbf{Methods}                        & \textbf{Value} & \textbf{Baidu-API} & \textbf{Alibaba-API} & \textbf{iFlytek-API} \\ \hline
\multirow{3}{*}{\textbf{Down-sampling}} & 5200           & 0/20               & 0/20                 & 0/20                 \\ \cline{2-5} 
                                        & 5600           & 0/20               & 0/20                 & 1/20                 \\ \cline{2-5} 
                                        & 6000           & 1/20               & 2/20                 & 2/20                 \\ \hline
\multirow{3}{*}{\textbf{Adding noise}}  & 500            & 2/20               & 2/20                 & 2/20                 \\ \cline{2-5} 
                                        & 1000           & 1/20               & 1/20                 & 1/20                 \\ \cline{2-5} 
                                        & 2000           & 0/20               & 1/20                 & 0/20                 \\ \hline
\end{tabular}
\end{adjustbox}
\end{table}

\subsection*{\textbf{ Limitations}}

In this study, we employed three speech recognition models as base models for the ensemble adversarial attack. However, we did not further explore the impact of the number of base models on the training of adversarial examples due to the limited availability of base models. Besides, the choice of the carrier is crucial for the success of the adversarial examples. In particular, a single music element is more susceptible to successful attacks with fewer perturbations than a multi-element music carrier. However, the specific audio features underlying this phenomenon are still unclear. In future research, we plan to increase the number of base models and investigate the inherent relationships between multiple models. We will also explore algorithms for picking appropriate carriers for adversarial examples to improve the robustness and interpretability of deep learning algorithms.

\section*{Conclusion}

Our study investigates the potential factors that influence the transferability of audio AEs. We reveal the subtle role of noise in the attack and the transferability of these examples. Specifically, different levels of noise can promote or restrict the transferability of AEs. Furthermore, we find that scale invariance and the presence of silence frames are positively correlated with transferability. Regarding the carriers used to generate AEs, we find that dialogue carriers provide better concealment and have a higher success rate than rock music carriers. This indirectly confirms the potential positive effect of silence frames because dialogue carriers contain many silent segments. Overall, our study sheds light on the various factors that impact the transferability of audio adversarial examples and provides insights into the development of more robust speech recognition systems.
Through numerous phoneme analyses, we believe that the decision boundaries of deep learning-based speech recognition models are irregular, which poses a challenge for transferring AEs. But we believe there exists an optimal attack path that can increase the success rate of attacks while reducing auditory perception between these models. Through experiments, we have observed that ensemble methods can enhance the transferability of adversarial examples. So we propose two ensemble transfer attack methods: RGE and DGWE, which obtained excellent attack success rates on iFlytek, Tencent, and Baidu APIs. We also note that the $p$ value in the self-integration is a crucial parameter that can influence the transferability of the AEs. 



\begin{backmatter}
\section*{Abbreviations}
\makebox[1.5cm][l]{DL}Deep learning \\
\makebox[1.5cm][l]{AEs}Adversarial Examples \\
\makebox[1.5cm][l]{DNNs}Deep Neural Networks \\
\makebox[1.5cm][l]{API}Application Programming Interface \\
\makebox[1.5cm][l]{RGE}Random Gradient Ensemble \\
\makebox[1.5cm][l]{DGWE}Dynamic Gradient Weighting Ensemble \\
\makebox[1.5cm][l]{TR}Transfer Rate \\
\makebox[1.5cm][l]{DS}The Deepspeech Attack \\
\makebox[1.5cm][l]{IRTA}The attack method in ``Imperceptible, robust, and targeted adversarial examples for automatic speech recognition" \\
\makebox[1.5cm][l]{ASR}Automatic Speech Recognition \\
\makebox[1.5cm][l]{CS}The CommanderSong Attack \\
\makebox[1.5cm][l]{DW}The Devil’s Whisper Attack \\
\makebox[1.5cm][l]{DC}The Disappeared Command Attack \\

\section*{Acknowledgements}
Not applicable.



\section*{Funding}
The authors are supported in part by NSFC No.62202275 and Shandong-SF No.ZR2022QF012 projects.

\section*{Availability of data and materials}
The code is available at: xxxx

\bibliographystyle{bmc-mathphys} 
\bibliography{refer}      





\end{backmatter}
\end{document}